\documentclass[a4paper]{jpconf}
\usepackage{graphicx}
\begin{document}
\title{Initial performance of the GlueX DIRC detector}

\author{A~Ali$^2$, F~Barbosa$^4$, J~Bessuille$^5$, E~Chudakov$^4$, R~Dzhygadlo$^2$, C~Fanelli$^{4,5}$, J~Frye$^3$, J~Hardin$^5$, A~Hurley$^7$, E~Ihloff$^5$, G~Kalicy$^1$, J~Kelsey$^5$, W~B~Li$^7$, M~Patsyuk$^6$, J~Schwiening$^2$, M~Shepherd$^3$, J~R~Stevens$^7$, T~Whitlatch$^4$, M~Williams$^5$ and Y~Yang$^5$}

\address{$^1$ Catholic University of America, Washington DC, U.S.A.}
\address{$^2$ GSI Helmholtzzentrum f\"ur Schwerionenforschung GmbH, Darmstadt, Germany}
\address{$^3$ Indiana University, Bloomington, IN, U.S.A.}
\address{$^4$ Jefferson Lab, Newport News, VA, U.S.A.}
\address{$^5$ Massachusetts Institute of Technology, Cambridge, MA, U.S.A.}
\address{$^6$ Joint Institute for Nuclear Research, Dubna, Russia}
\address{$^7$ William $\&$ Mary, Williamsburg, VA, U.S.A.}

\ead{r.dzhygadlo@gsi.de}

\begin{abstract}
  The GlueX experiment at Jefferson Laboratory aims to perform quantitative tests of non-perturbative QCD by studying the spectrum of light-quark mesons and baryons. A Detector of Internally Reflected Cherenkov light (DIRC) was installed to enhance the particle identification (PID) capability of the GlueX experiment by providing clean $\pi$/K separation up to 3.7 GeV/$c$ momentum in the forward region ($\theta<11^{\circ}$), which will allow the study of hybrid mesons decaying into kaon final states with significantly higher efficiency and purity. The new PID system is constructed with radiators from the decommissioned BaBar DIRC counter, combined with new compact photon cameras based on the SuperB FDIRC concept. The full system was successfully installed and commissioned with beam during 2019/2020. The initial PID performance of the system was evaluated and compared to one from Geant4 simulation.
\end{abstract}

\section{Introduction}

\begin{figure}[h]
  \centering
  \includegraphics[width=0.8\columnwidth]{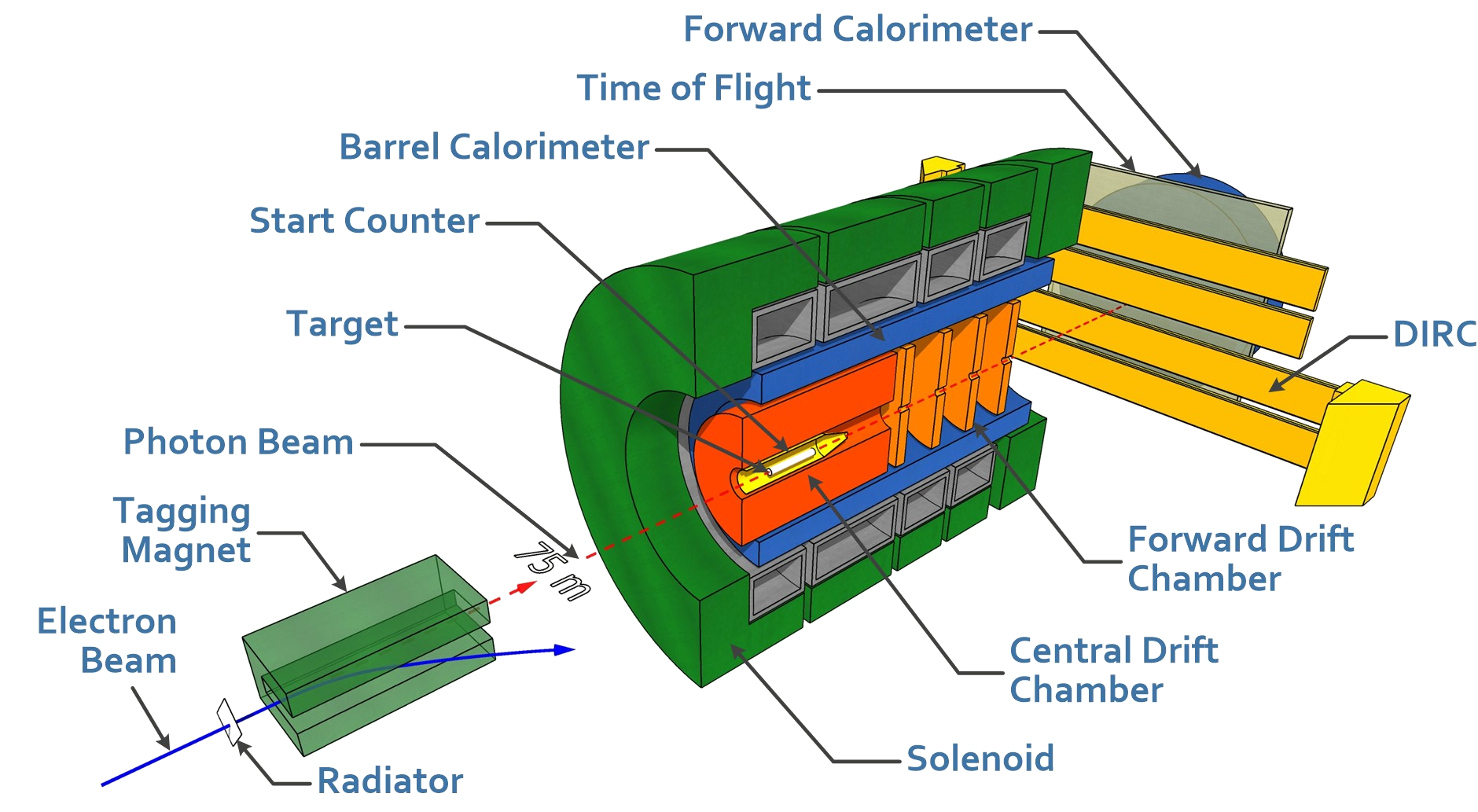}
  \caption{\label{gluex_det}  Schematic of the Hall D beamline and GlueX detector at Jefferson Laboratory. The DIRC detector is installed directly upstream of the time-of-flight detector in the forward region.}
\end{figure}

The GlueX experiment (see figure \ref{gluex_det}), located in Hall D at Jefferson Laboratory, uses a tagged photon beam to study the spectrum of light-quark mesons and baryons \cite{gluex1,gluex2,gluex3}. In the initial detector configuration the particle identification (PID) was provided by a time-of-light system and was limited to the tracks with momenta up to 2 GeV/$c$. For Phase II the physics program was extended to also include the high luminosity strangeness study which requires more advanced PID. Therefore, the DIRC (Detection of Internally Reflected Cherenkov light) detector was installed with the goal to extend the PID and to reach 3 standard deviations for $\pi$/K separation up to 3.7~GeV/$c$.


\section{Detector construction and installation}

The GlueX DIRC \cite{gluex_tdr,gluex_prog} reuses radiators from the decommissioned BaBar DIRC \cite{babar}. Four unmodified bar boxes, each containing twelve 4.9 m long fused silica radiator bars, are attached to two newly constructed compact photon expansion volumes based on the SuperB FDIRC concept \cite{fdirc}. The expansion volume is filled with distilled water to match refractive indices of the radiator and photon detectors.  A set of mirrors is used to focus the Cherenkov photons on an array of Hamamatsu H12700 Multi-Anode Photomultipliers (MaPMTs). Each MaPMT has 64 pixels with size of $6 \times 6$~mm$^{2}$ and together with MAROC-based electronics allows to detect photons with a time resolution of approximately 0.8 ns. In total 11520 readout channels are used.

During operation, the MaPMTs in each expansion volume are directly illuminated by LED sources to provide a reference signal for time calibration and to monitor the condition parameters of the detector system, including water quality and MaPMTs efficiency.

The complete GlueX DIRC detector was installed and commissioned during 2019. The first production data were collected in 2020 as a part of Phase II program. Collected data were used to determine the initial performance of the DIRC system. 

\section{Performance}

\begin{figure}[t]
  \vspace{-2mm}
  beam data
  \vspace{-2mm}
  \begin{center}
    \includegraphics[width=0.495\columnwidth]{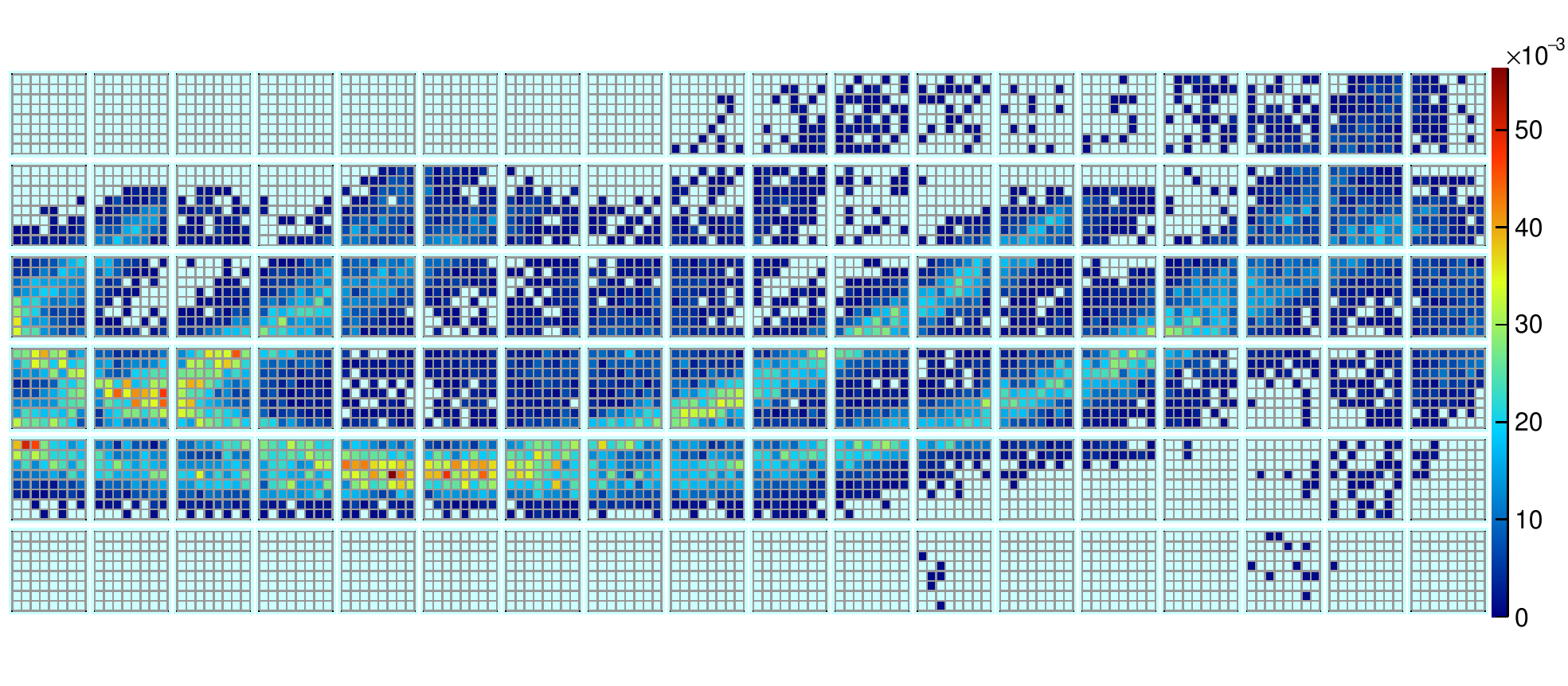}
    \includegraphics[width=0.495\columnwidth]{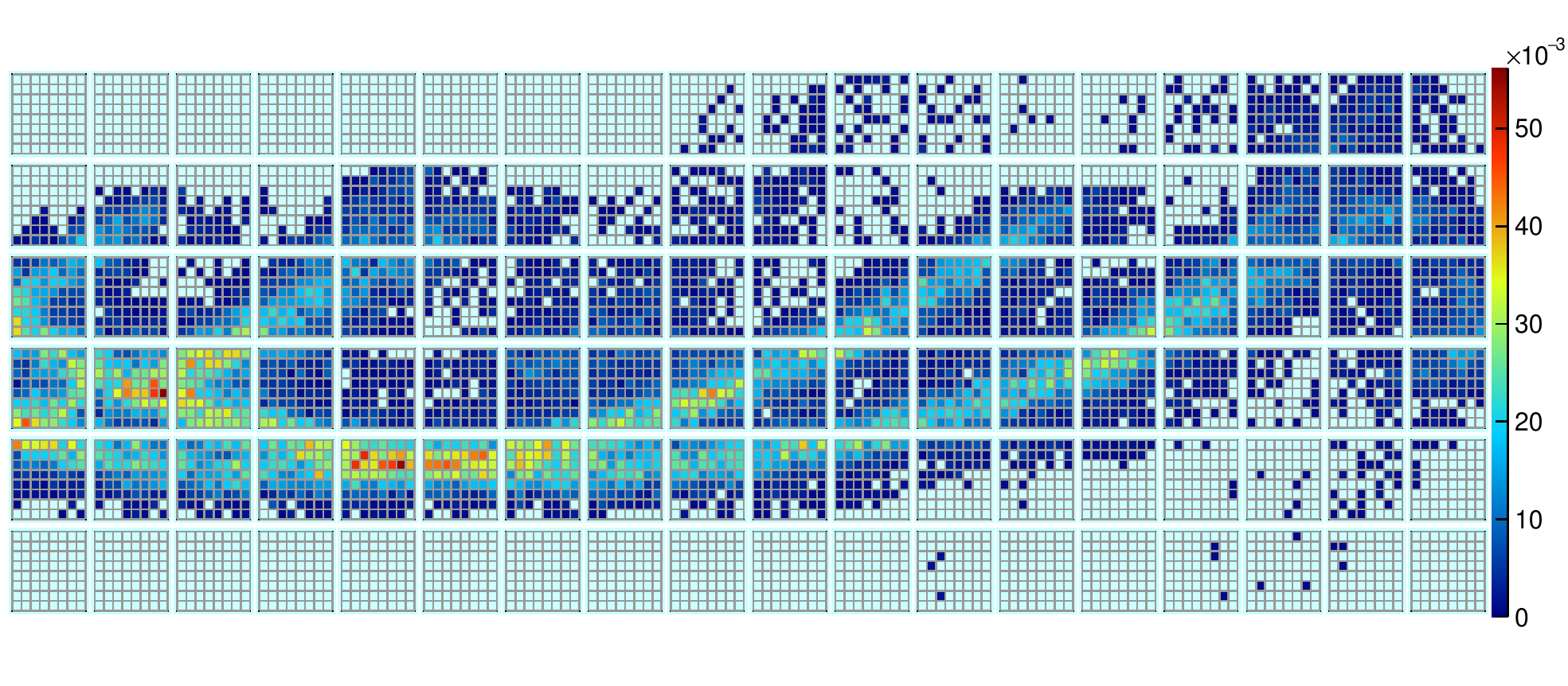}
  \end{center}
  \vspace{-2mm}
  simulation
  \vspace{-2mm}
  \begin{center}
    \includegraphics[width=0.495\columnwidth]{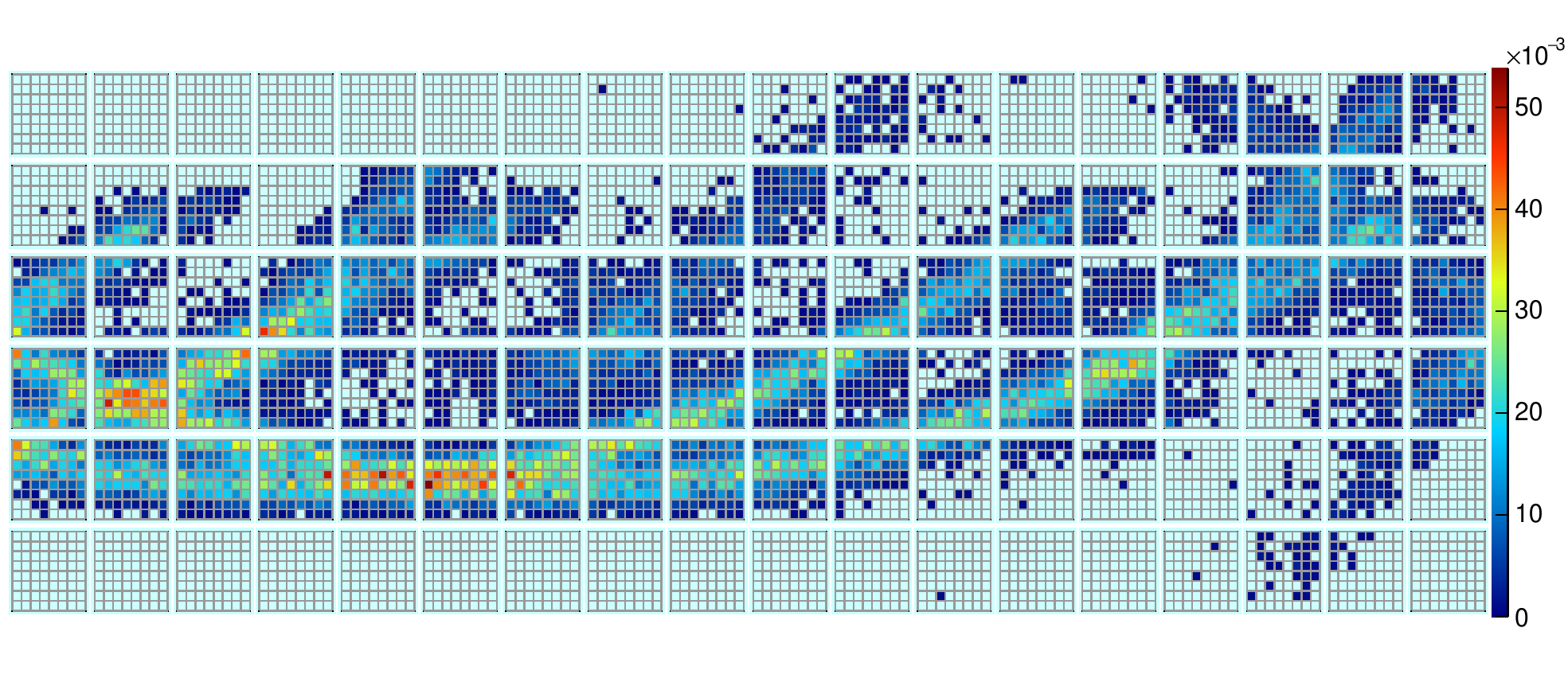}
    \includegraphics[width=0.495\columnwidth]{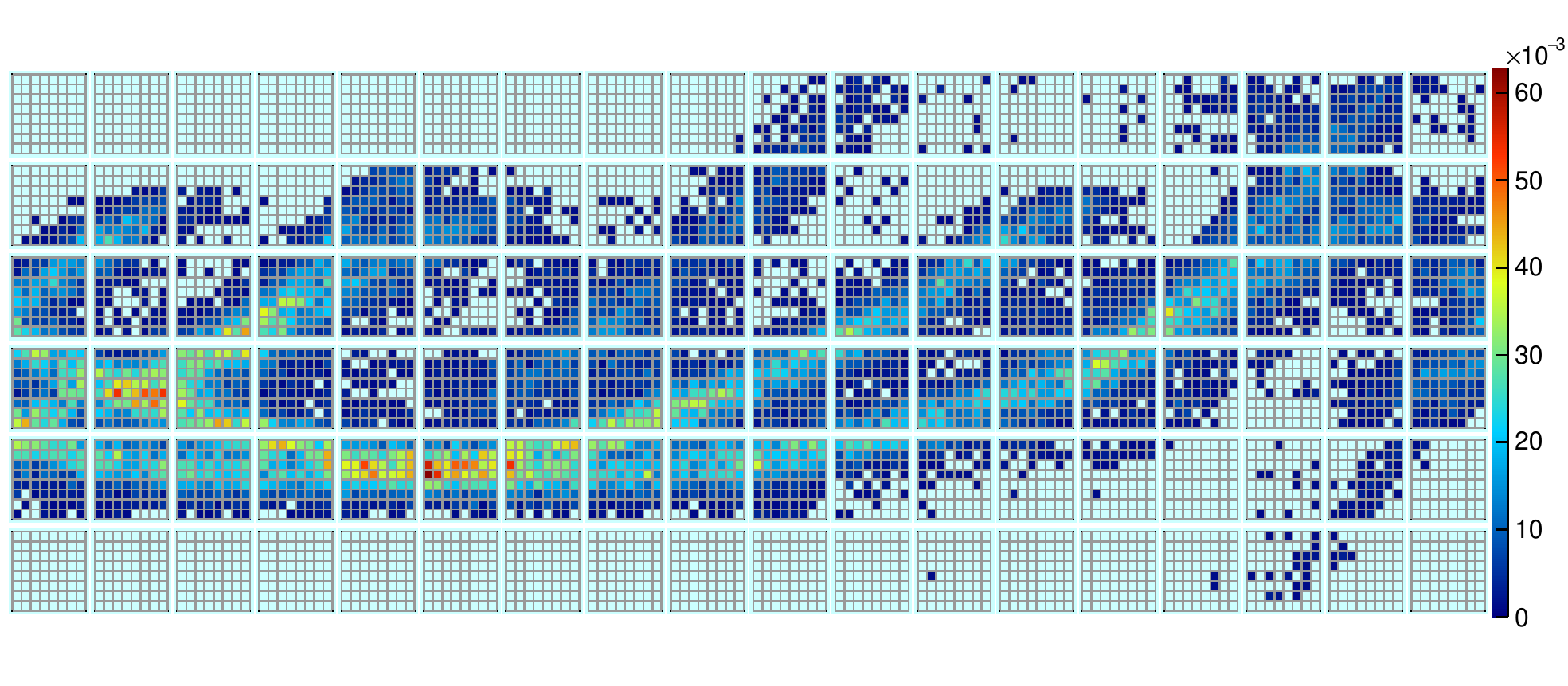}
  \end{center}

  \caption{\label{fig_hitpat} Accumulated hit pattern for 1000 pions (left) and kaons (right) at 3.5~GeV/$c$ momentum for beam data (top) and Geant4 simulations (bottom). The difference between pions and kaons represented with vertical shift of the pattern by one pixel.}
\end{figure}

The sample of pions and kaons, kinematically identified using exclusive $\gamma p \rightarrow \rho p$, $\rho \rightarrow \pi^{+}\pi^{-}$ and $\gamma p \rightarrow \phi p$, $\phi \rightarrow K^{+}K^{-}$ reactions, was used as an external PID for the GlueX DIRC performance evaluation. 
As the basic observables the GlueX DIRC provides the hit positions of the Cherenkov photons, the multiplicity of the detected photons, and the propagation time of those photons inside the DIRC. An example of accumulated hit pattern detected by the array of MaPMTs is shown in figure~\ref{fig_hitpat} for 1000 pions (left) and kaons (right) at 3.5 GeV/$c$ momentum flying in the same direction. The patterns from the beam data (top) are well described by the Geant4 MC simulations (bottom).

\begin{figure}[bht]
  \includegraphics[width=0.49\columnwidth]{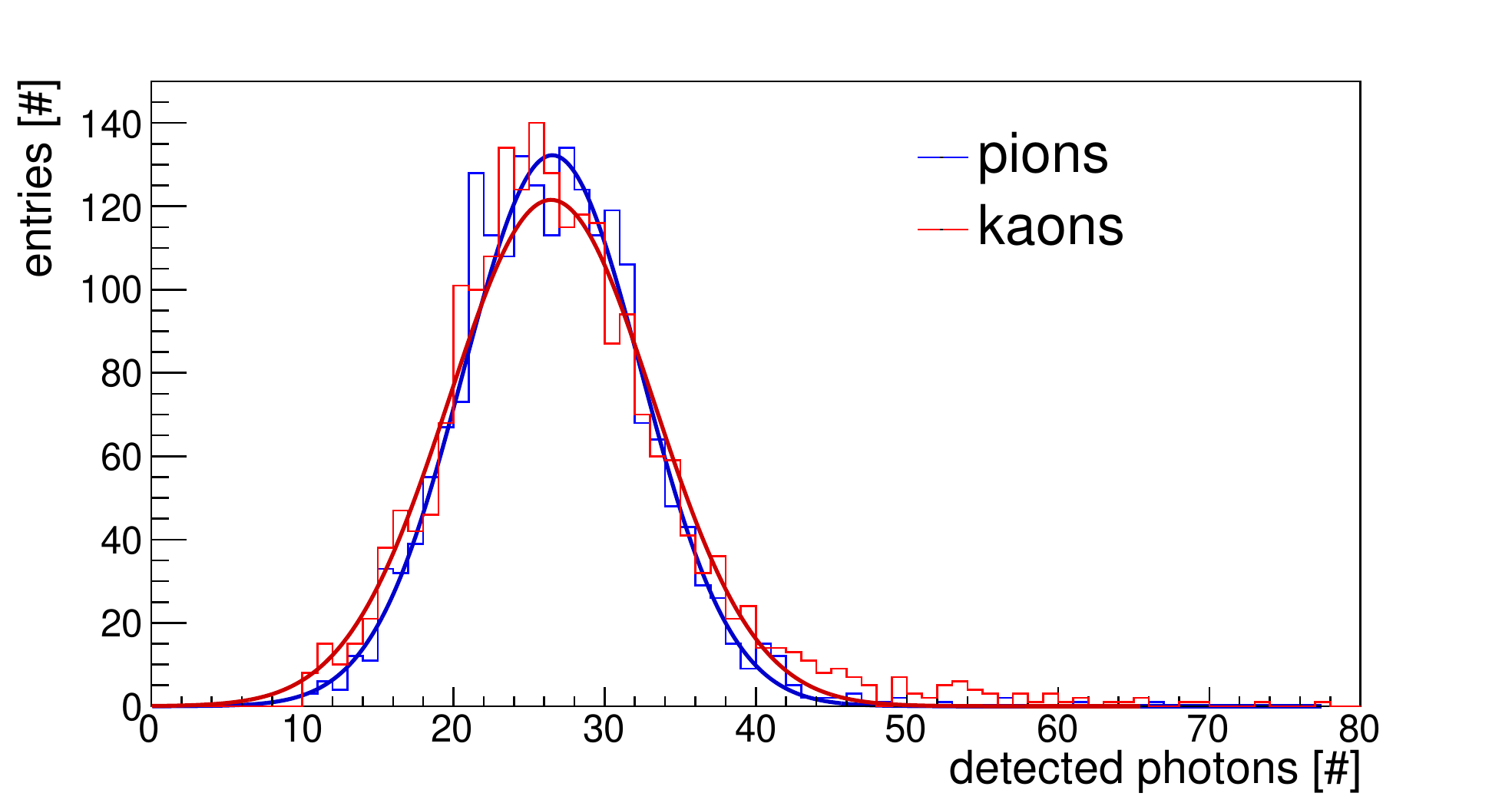}
  \includegraphics[width=0.49\columnwidth]{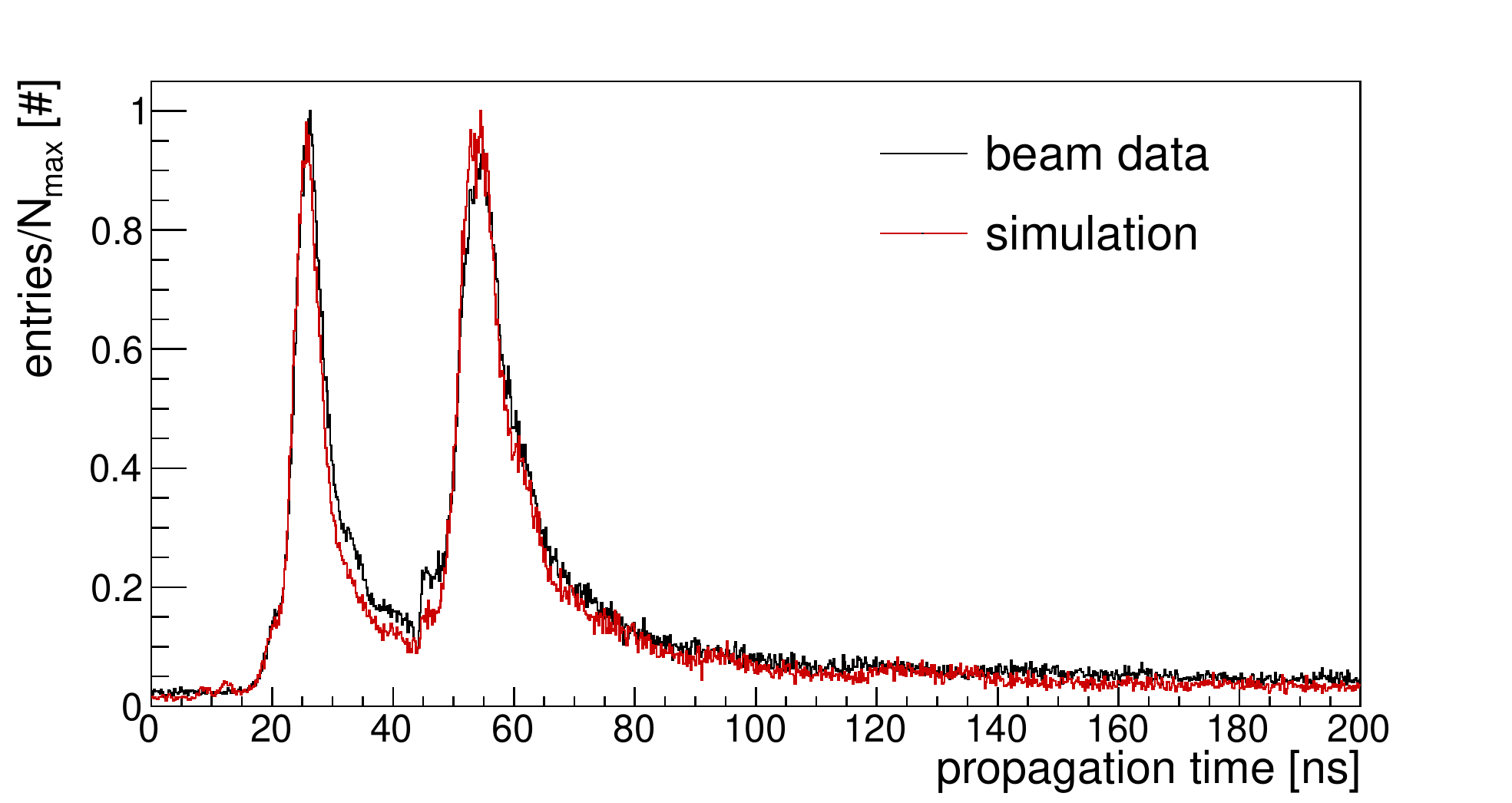}
  \caption{\label{fig_time} Detected photon yield for 2000 pions and kaons (left) and propagation time of those photons in comparison to Geant4 simulation (right).}
\end{figure}

The average multiplicity of the detected photons per track ranges from 10 to 30, depending on the hit position of the charged track at the radiator. Figure~\ref{fig_time} (left) shows the detected photon yield distribution for pions and kaons at 3.5~GeV/$c$ momentum hitting the central region of the radiator. A Gaussian fit describes both distributions well and gives a mean of 27 photons which is about 35\% less then the predicted yield from Geant4 simulation. The unexpected loss of photons happens as a result of reflectivity degradation of the aluminized mirrors which were exposed to the water inside of the optical box. The protective layer of borosilicate glass on top for the mirror surfaces is foreseen for the next data taking periods \cite{gluex_wenliang}.
The measured propagation time of the Cherenkov photons is in a good agreement with simulation as shown by figure~\ref{fig_time} (left). The first and the second peaks in time spectra are corresponding to the forward and backward going photons, respectively.

\begin{figure}[h]
  \includegraphics[width=0.49\columnwidth]{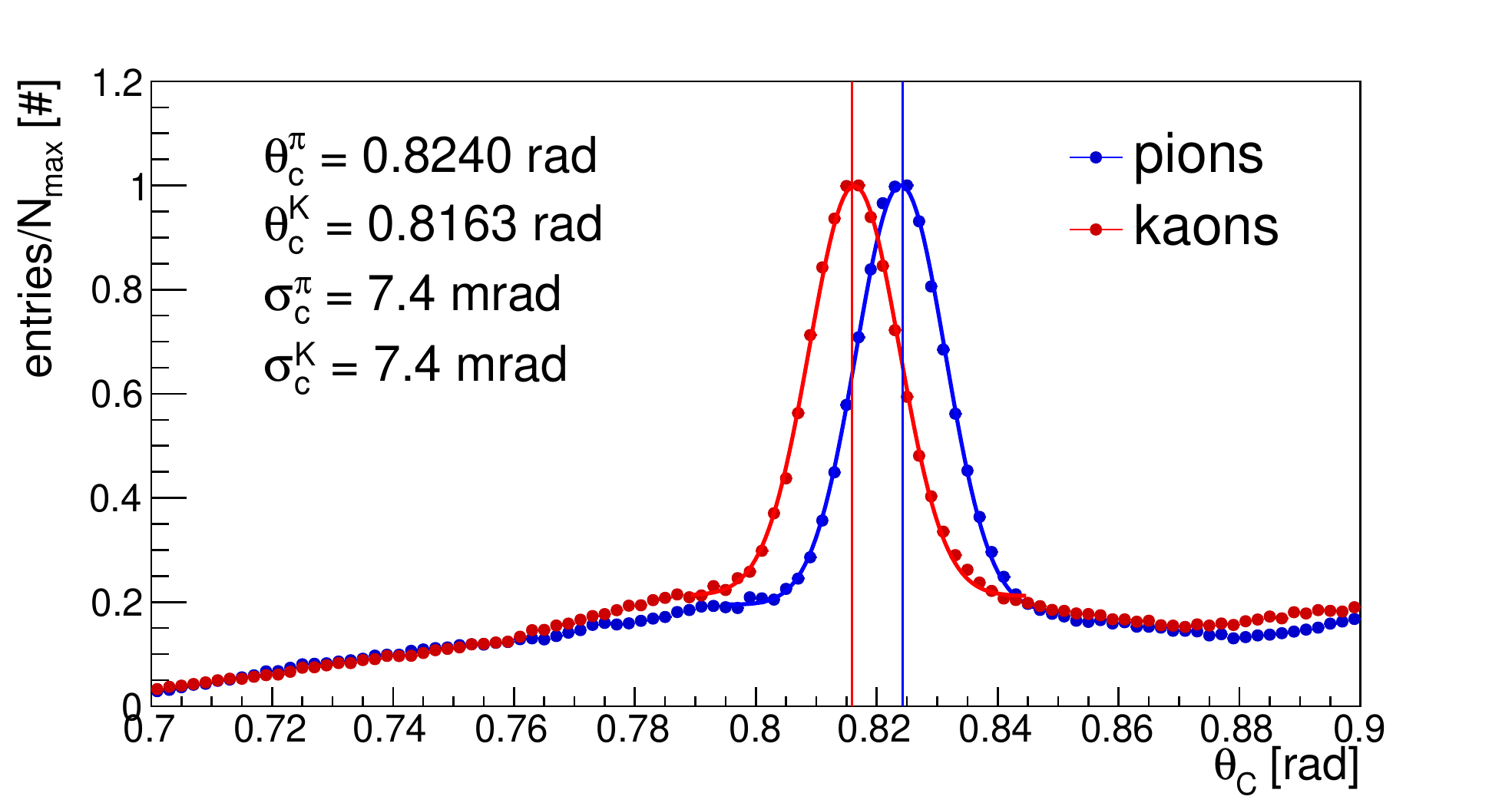}
  \includegraphics[width=0.49\columnwidth]{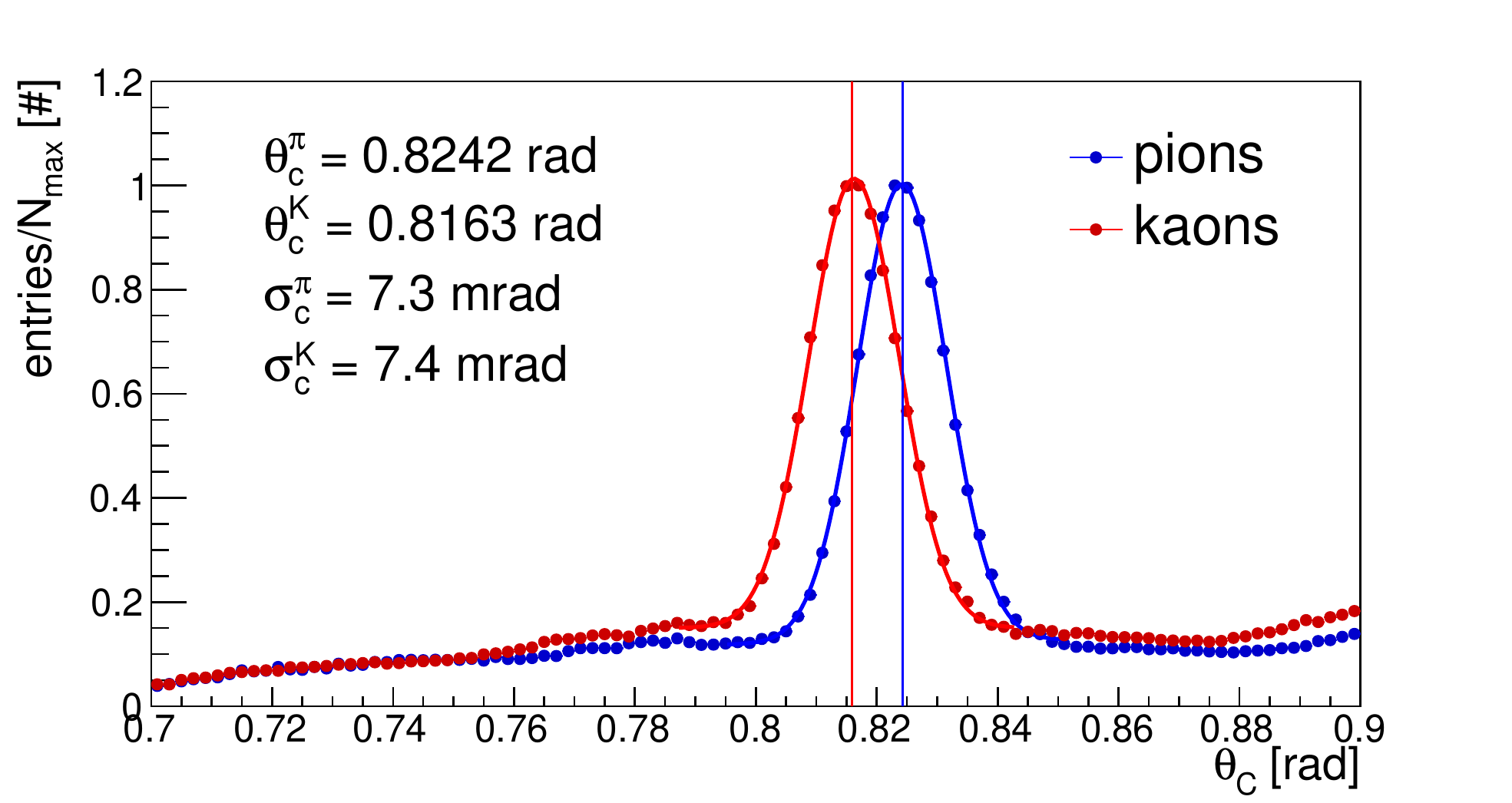}
  \caption{\label{fig_cangle} Single photon Cherenkov angle distribution for 2000 $\pi$/K at 3.5 GeV/$c$ momentum from beam data (left) and Geant4 simulation (right). The vertical lines shows the expected values of Cherenkov angle.}
\end{figure}

\begin{figure}[bht]
  \centering
  \includegraphics[width=0.33\columnwidth]{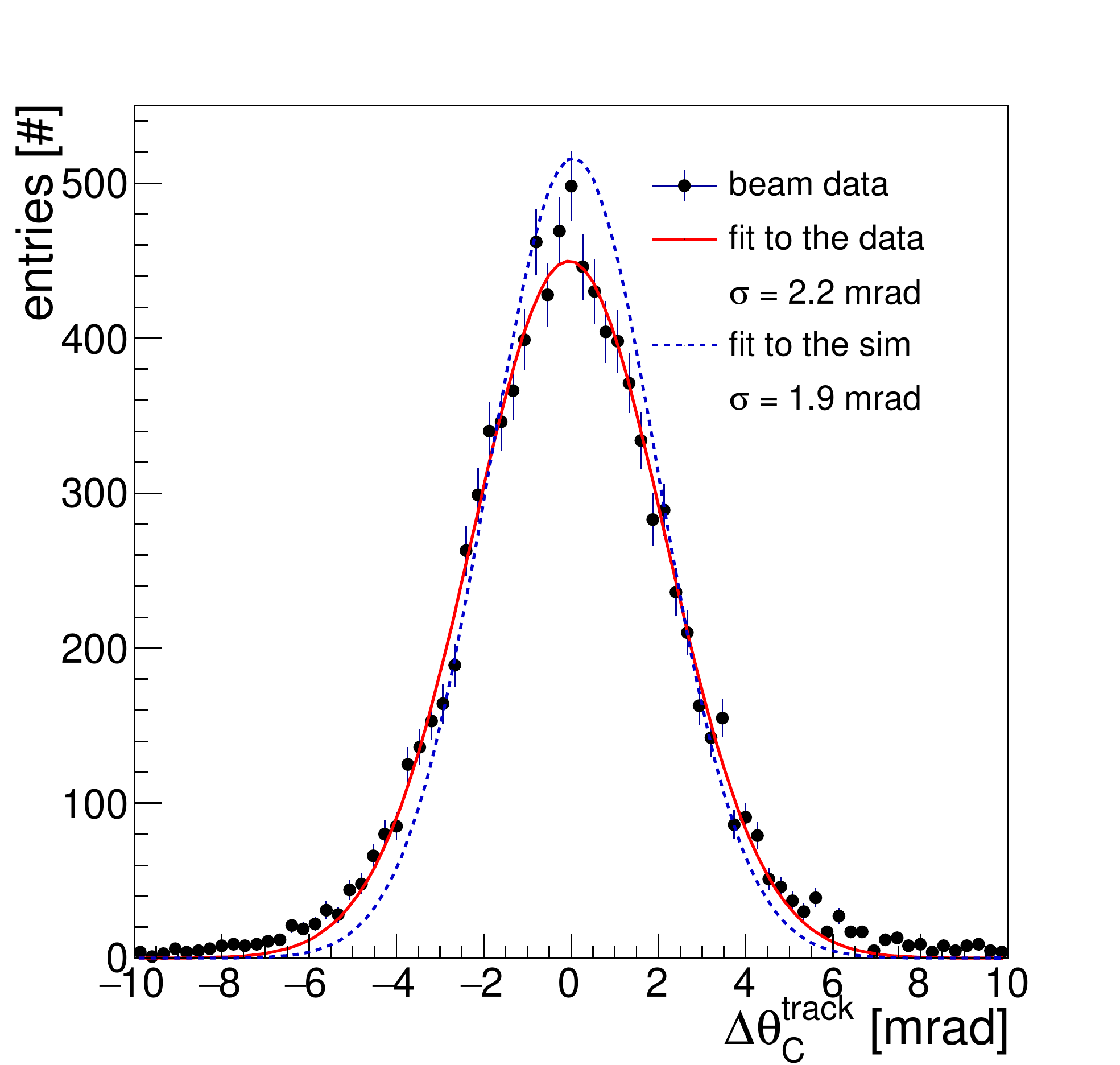}
  \includegraphics[width=0.59\columnwidth]{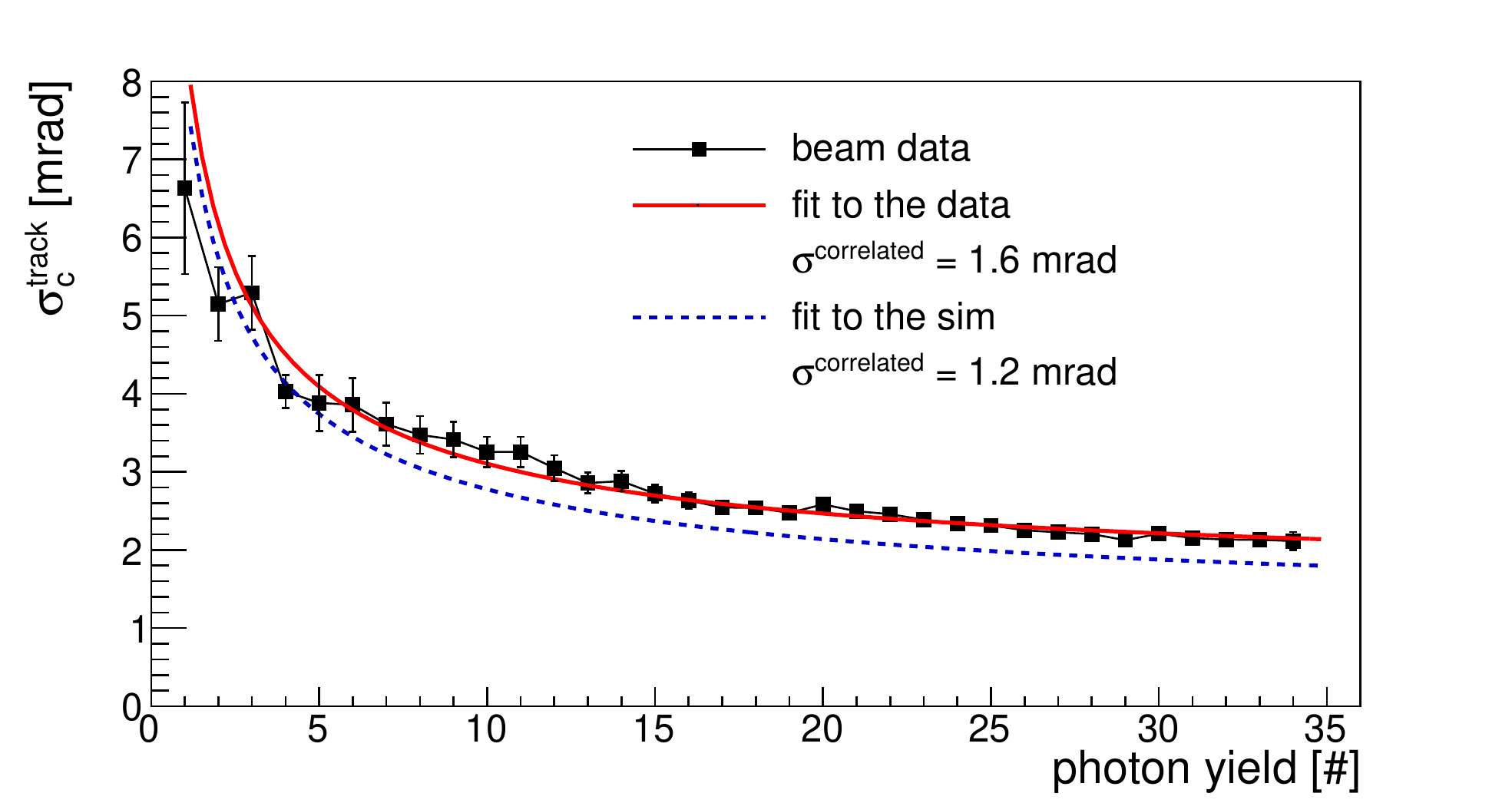}
  \caption{\label{fig_track} Cherenkov angle difference between reconstructed and expected values for 10k pions at 3.5 GeV/$c$ (left) and Cherenkov angle resolution evaluated for different photon yield (right).}
\end{figure}

The \emph{geometrical reconstruction} method \cite{barrel_reco} was used to evaluate the PID performance of the DIRC. The method uses pre-simulated look-up tables (LUTs) of all possible exit vectors of the photons at the end of the radiator which can lead to a hit in MaPMT's pixel. The Cherenkov angle is determined by combining the charged particle's direction vector with all possible photon directions from the LUT for a given pixel. Figure~\ref{fig_cangle} shows the reconstructed Cherenkov angle distribution for pions and kaons at 3.5 GeV/$c$ momentum from beam data (left) and Geant4 simulation (right).
The distributions are fitted with a sum of the Gaussian and polynomial function. The polynomial part accounts for the background which arises due to ambiguous photons path inside the expansion volume. The expected values of the Cherenkov angle, shown with vertical lines, are in excellent agreement with the means of the Gaussian fit. The single photon resolution $\sigma_{\mathrm{C}} = 7.4\pm0.1$~mrad was obtained for both beam data and Geant4 simulations. The achieved agreement is a result of multiple corrections including per-PMT Cherenkov angle and chromatic corrections.
The single photon resolution allows to disentangle different contributions to the Cherenkov angle resolution per track.  The latter directly represents the PID performance of the DIRC and can be written as $\mathrm{(\sigma_{C}^{track})^2 = \sigma^2_{C}/N_{photons}+(\sigma^{correlated})^2}$, where $N_{\mathrm{photons}}$ is the number of detected photons and $\sigma^{\mathrm{correlated}}$ represents the tracking resolution and any possible misalignment of the geometry. The $\mathrm{\sigma_{C}^{track}}$ can also be evaluated directly by plotting the difference between reconstructed and expected Cherenkov angle on track-by-track basis, as shown on figure~\ref{fig_track} (left). The width of the Gaussian fit $\mathrm{\sigma_{C}^{track}} = 2.2\pm0.1$ mrad for beam data and $1.9\pm0.1$ mrad for simulation. The $\pi$/K separation then can be estimated as the Cherenkov angle difference between $\pi$ and K (8.5 mrad at 3.5 GeV/$c$) divided by $\mathrm{\sigma_{C}^{track}}$, resulting in 3.8~$\sigma$ for beam data, although the non-Gaussian tail in the distribution will decrease the PID performance.
Evaluating the Cherenkov angle resolution per track for different photon yield allows to extract the correlated term (see figure~\ref{fig_track}, right). The obtained value of 1.6 mrad and 1.2 mrad for beam data and Geant4 simulation, respectively. The larger $\sigma^{\mathrm{correlated}}$ of the beam data indicates that further calibration improvements are possible. 

\begin{figure}[t]
  \includegraphics[width=0.49\columnwidth]{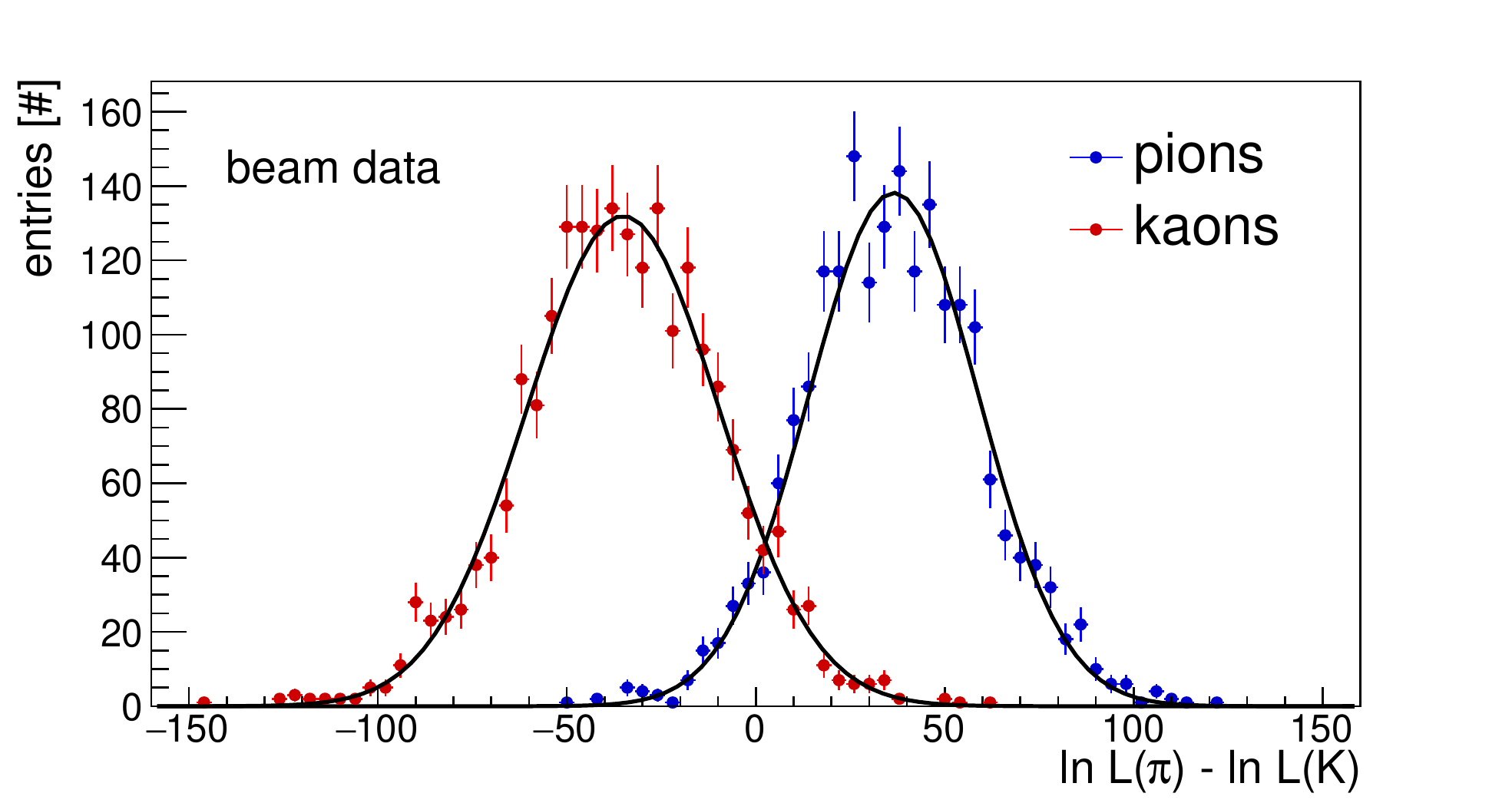}
  \includegraphics[width=0.49\columnwidth]{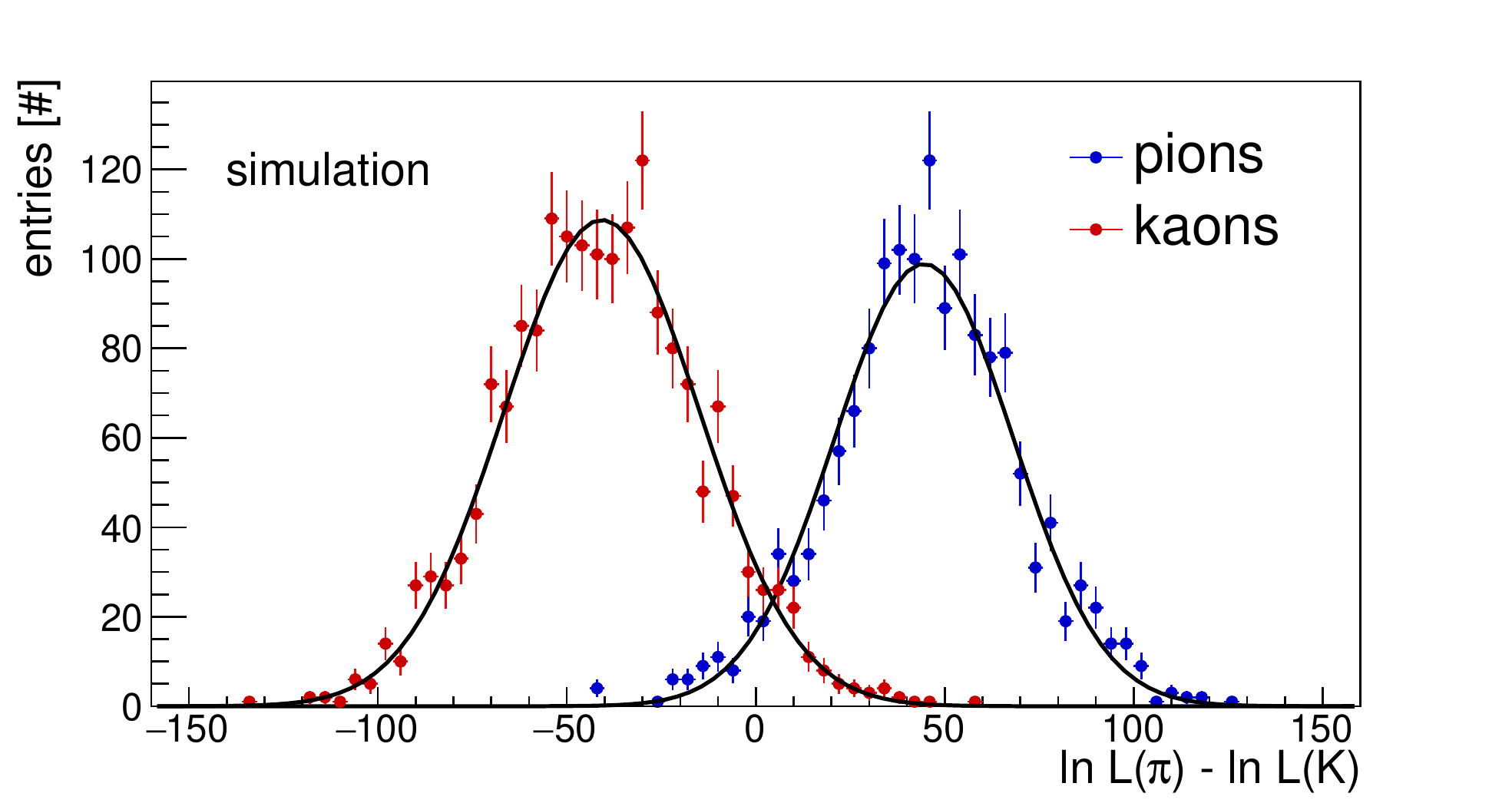}
  \caption{\label{fig_log} Pion-kaon log-lilkelihood difference distribution for 4000 events from beam data (left) and Geant4 simulation (right). The obtained separation power is  $3.0\pm0.1$ and $3.4\pm0.1$ for beam data and simulation, respectively. }
\end{figure}

Particle identification is done by performing a track-by-track unbinned likelihood calculation of the reconstructed Cherenkov angle distribution to different particle hypothese. Figure~\ref{fig_log} shows the $\pi$/K log-lilkelihood difference distribution for 3.5 GeV/$c$ momentum particles. The separation of $3.0\pm0.1$ $\sigma$ and $3.4\pm0.1$ $\sigma$ is achieved for the beam data and simulation, respectively. The smaller separation for the beam data reflects the differences in the correlated term.

\section{Conclusion}
The GlueX DIRC was installed and successfully commissioning during February and December, 2019. The first production data were collected in 2020 as a part of Phase II program. The initial data analysis show very good agreement of the hit patterns and propagation times of the Cherenkov photons with expected distribution from Geant4 simulations. Furthermore, the reconstructed Cherenkov angle resolution per track agrees well with design values. The $\pi$/K separation of 3 s.d. is confirmed for 3.5 GeV/$c$ momentum tracks. 

\ack
We would like to acknowledge the outstanding efforts of the staff of the Accelerator and the Physics Divisions at Jefferson Lab that made the experiment possible.  This work is supported by the U.S. Department of Energy, Office of Science, Office of Nuclear Physics under contracts DE-AC05-06OR23177, DE-FG02-05ER41374 and Early Career Award contract DE-SC0018224 and the German Research Foundation, GSI Helmholtzzentrum f\"ur Schwerionenforschung GmbH.


\section*{References}
\begin{thebibliography}{9}

\bibitem{gluex1} GlueX Collaboration, Mapping the spectrum of light quark mesons and gluonic excitations withlinearly polarized protons, Jefferson Lab PAC 30 proposal, U.S.A. (2006)

\bibitem{gluex2} Sejevs A et al. (GlueX Collaboration), An initial study of mesons and baryons containing strange quarks with GlueX ,Jefferson Lab PAC 40 proposal, U.S.A. (2013) [arXiv:1305.1523]

\bibitem{gluex3} Dugger M et al. (GlueX Collaboration), A study of decays to strange final states with GlueX in Hall D using componentsof the BaBar DIRC, Jefferson Lab PAC 42 proposal, U.S.A. (2014) [arXiv:1408.0215]

\bibitem{gluex_tdr} GlueX Collaboration, GlueX DIRC Technical Design Report (2015)

\bibitem{gluex_prog} Ali A et al. The GlueX DIRC program, {\it JINST} 15 (2020) {\bf 09}, C04054
    
\bibitem{babar} Adam I et al., The DIRC particle identification system for the BaBar experiment, \emph{Nucl. Instr. and Meth. Res. Sect} {\bf A538} (2005) 281
  
\bibitem{fdirc} Dey B et al., Design and performance of the Focusing DIRC detector, {\it Nucl. Instrum. Meth.} A {\bf 775} (2015) 112
  
\bibitem{gluex_wenliang} Ali A et al. Installation and Commissioning of the GlueX DIRC, {\it JINST} 15 (2020) {\bf 09}, C09010

\bibitem{barrel_reco} Dzhygadlo R et al., Simulation and reconstruction of the PANDA BarrelDIRC, \emph{Nucl. Instr. and Meth. Res. Sect} {\bf A766} (2014) 263
  
\end{thebibliography}
\end{document}